\documentclass[iop, numberedappendix]{emulateapj}
\usepackage{apjfonts}

\usepackage{color}
\definecolor{citeRGB}{rgb}{0,0.1,0.7}
\usepackage[hyperfootnotes=true,naturalnames=true,letterpaper,pdfstartview=FitH,pdfpagemode=UseNone,colorlinks=true,citecolor=citeRGB]{hyperref}

\gdef\HST{\textit{HST}}
\gdef\G141{\textit{G141}}
\gdef\F140W{\textit{F140W}}
\gdef\fluxcgs{\mathrm{erg\ s^{-1}\ cm^{-2}}}

\gdef\flux_radius{\textsc{flux\_radius}}
\gdef\epers{\textit{e}$^{-}$ s$^{-1}$}

\gdef\kms{km\,s$^{-1}$}
\gdef\mum{\mu\mathrm{m}}
\gdef\24mum{$24\,\mu\mathrm{m}$}
\gdef\arcsec{^{\prime\prime}}

\gdef\4ang{4000\,\AA}

\gdef\UDFj{UDFj-39546284}
\gdef\compareID{UDF-40106456}

\gdef\othree{[\ion{O}{3}]$\lambda$4959,5007}
\gdef\epers{electrons~s$^{-1}$}

\gdef\lineflux{$3.5\pm1.3\times10^{-18}\,\fluxcgs$}

\shortauthors{Brammer et al.}
\shorttitle{\HST\ spectroscopy of UDFj-39546284}
\slugcomment{ApJL, accepted}

\begin{document}

\title{A tentative detection of an emission line at $1.6\,\mum$ for the $z\sim12$ candidate UDFj-39546284\footnotemark[0]}

\footnotetext[0]{Based on observations made with the NASA/ESA \textit{Hubble
Space Telescope}, programs GO-12099, 12177 and 12547, obtained at the
Space Telescope Science Institute, which is operated by the Association of
Universities for Research in Astronomy, Inc., under NASA contract NAS
5-26555.}

\author{Gabriel B. Brammer\altaffilmark{1},
Pieter G.\ van Dokkum\altaffilmark{2},
Garth D.\ Illingworth\altaffilmark{4},
Rychard J.\ Bouwens\altaffilmark{3},
Ivo Labb\'e\altaffilmark{3}, 
Marijn Franx\altaffilmark{3},
Ivelina Momcheva\altaffilmark{2},
Pascal A.\ Oesch\altaffilmark{4}}


\email{gbrammer@eso.org}

\altaffiltext{1}
{European Southern Observatory, Alonso de C\'ordova 3107, Casilla 19001, Vitacura, Santiago, Chile}
\altaffiltext{2}
{Department of Astronomy, Yale University, New Haven, CT 06520, USA}
\altaffiltext{3}
{Leiden Observatory, Leiden University, NL-2300 RA Leiden, Netherlands}
\altaffiltext{4}
{UCO/Lick Observatory, University of California, Santa Cruz, CA 95064}

\begin{abstract}

We present deep WFC3 grism observations of the candidate $z\sim12$ galaxy
\UDFj\ in the \HST\ Ultra Deep Field (UDF), by combining spectroscopic data
from the 3D-HST and CANDELS surveys. The total exposure time is 40.5~ks and
the spectrum covers $1.10 < \lambda < 1.65\,\mum$. We search for faint
emission lines by cross-correlating the 2D
G141 spectrum with the observed $H_{160}$ morphology, a technique that is unique to slitless spectroscopy at \HST\
resolution. We find a 2.7$\sigma$ detection of an emission line at
1.599~$\mum$---just redward of the $JH_{140}$ filter---with flux
\lineflux. Assuming the line is real, it contributes $110\pm40$\% of the
observed $H_{160}$ flux and has an observed equivalent width $> 7300$\,\AA.  If the line is confirmed, it could be Ly-$\alpha$ at $z=12.12$.  However, a more plausible interpretation, given current results, could be a lower redshift feature such as \othree\ at $z=2.19$.  We find
two other 3D-HST [\ion{O}{3}] emitters within 1000\,\kms\ of that redshift in the GOODS-South field.  Additional support for this interpretation comes from the discovery of a bright
``[\ion{O}{3}]~blob'' with a secure G141 grism
redshift of $z = 1.605$.  This object has a strikingly large observed equivalent width of nearly 9000\,\AA\ that results in similar ``dropout'' colors as
\UDFj.
    
\end{abstract}


\section{Introduction}
\label{s:introduction}

Over the past few years deep imaging programs with the WFC3 camera on the
\textit{Hubble Space Telescope} (\HST) have opened up the Universe at
$z > 8$ for systematic study \citep[e.g., ][]{bouwens:11, bouwens:12b, oesch:12a, ellis:13, coe:13}. 
Photometric data can provide strong constraints on the redshifts and
properties of very high redshift galaxies \citep[e.g., ][]{bouwens:11b, bouwens:12a}.
However, the interpretation is hampered by the lack of spectroscopic
information that provides unambiguous redshift determination and
the contributions of emission lines to the broad band fluxes. The equivalent
widths of rest-frame optical lines increase with redshift
\citep[e.g., ][]{fumagalli:12}, and the emission line contribution can be
significant even in the broad IRAC filters \citep[e.g.,][]{labbe:10, ono:10}.

Obtaining high quality spectra of these very distant objects is a formidable
challenge because of their faintness
and because key emission and absorption
features are redshifted into the observer's near-IR. 
Apart from a gamma ray burst at $z=8.1$ \citep{salvaterra:09, tanvir:09} the
only claimed \citep[and disputed;][]{bunker:13}
detection is a Ly-$\alpha$ line at $z=8.6$ in a
\textit{z}-dropout galaxy in the UDF \citep{lehnert:10}.

Here we describe a deep WFC3 G141 grism spectrum of one of the most
intriguing high redshift candidates yet discovered: \UDFj. This object was
identified as the first robust $z\gtrsim 10$ candidate by \citet{bouwens:11}: it was only detected in $H_{160}$ in the HUDF09 dataset \citep[see also][]{oesch:12a}.
Recent observations from the HUDF12 program \citep{ellis:13}
confirm the $H_{160}$ detection and show that it is undetected in $JH_{140}$
\citep{ellis:13, bouwens:13}. The $>1.4$ mag dropout (2$\sigma$) in the $JH_{140}$ filter
pushes its likely redshift to $z\sim12$, although alternative explanations
cannot be ruled out \citep{ellis:13}. Here we report a 2.7$\sigma$ detection of
an emission line for \UDFj\ that would account for most or all of its $H_{160}$
flux. Though this line could be Ly-$\alpha$ at $z=12$, we argue that it
is more likely to be \othree\ at $z=2.19$ based in part on the discovery of a
bright low-redshift analog with one of the highest [\ion{O}{3}] equivalent
widths ever observed (8700\,\AA) and broad-band colors similar to \UDFj. Magnitudes are in the AB system throughout.
 
\section{Observations \& Analysis}
\label{s:observations}

\begin{figure}
\epsscale{1.1}
\plotone{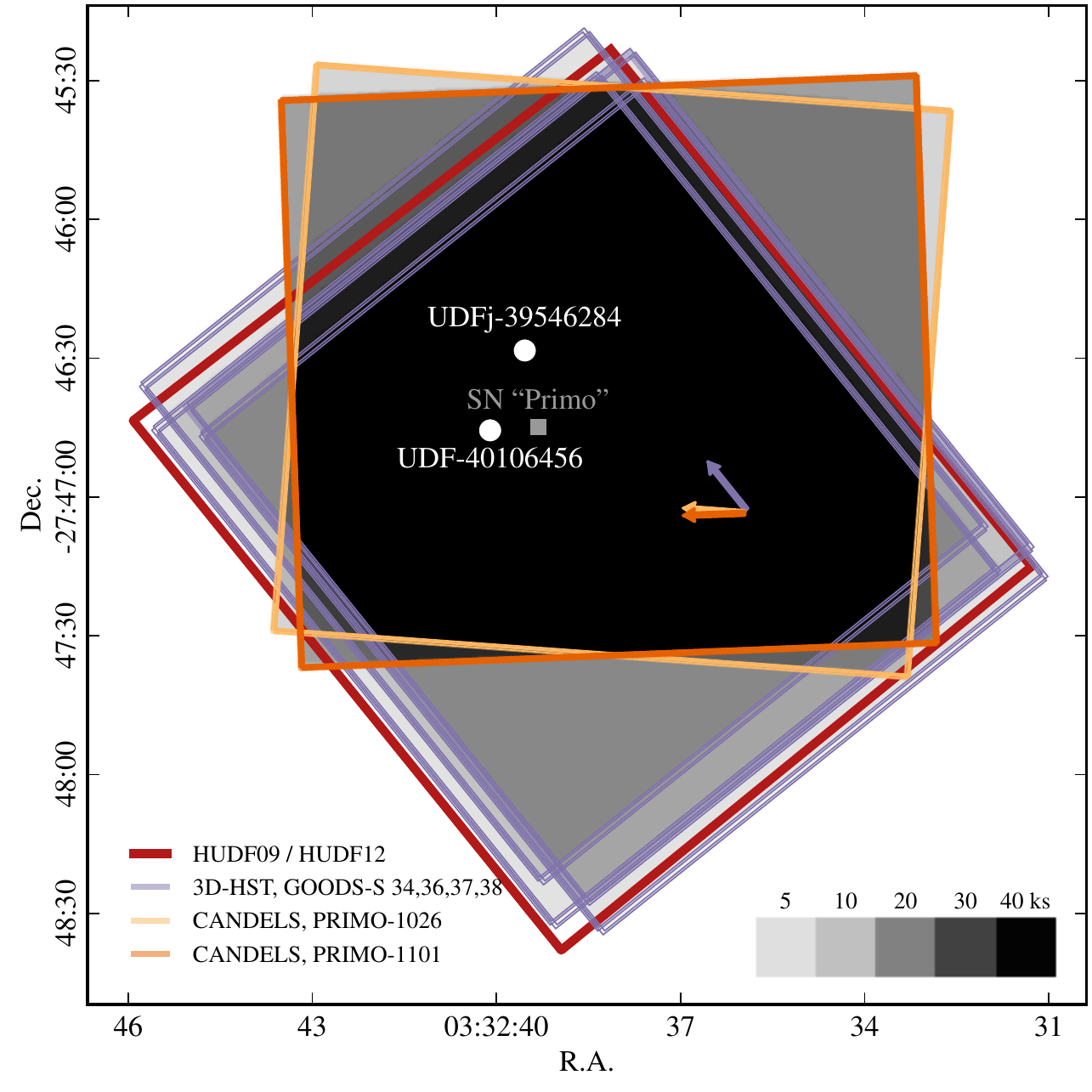} 
\caption{Exposure map of the 3D-HST and CANDELS G141 grism pointings in
the HUDF. The vectors indicate the direction and extent of the first-order G141
spectra. The locations of the CANDELS supernova and the two faint UDF galaxies described below are
indicated. These objects lie in the area of deepest G141
grism coverage, with a total exposure time of 40.5~ks (17
orbits).\label{f:expmap}}  
\end{figure}

The UDF was observed with the WFC3/G141 grism as part
of the 3D-HST survey \citep[GO-12177; PI: van Dokkum;][]{brammer:12a}. The
observations comprise four separate visits at the same orientation with G141 integrations of 4.7\,ks
(two orbits) each.  The UDF was also observed by the CANDELS supernova
follow-up program (GO-12099, PI: Riess) when a supernova at $z=1.55$,
nicknamed ``Primo'', was discovered in the first epoch of CANDELS imaging in the GOODS-South field\citep[see][]{rodney:12}.
The field was observed in
two visits on 2010 Oct 26 and 2010 Nov 01 with G141 integrations of 6.6 and 15 ks, respectively.  The
position angle of the grism dispersion axis is rotated by 6.4 degrees between
the two visits, which together are rotated 50 degrees with respect to the 3D-HST visits. The relation of the G141 pointings to the deep HUDF09+HUDF12
imaging area is shown in Fig.\,\ref{f:expmap}. 



The Primo and 3D-HST visits were pre-processed separately and identically
using the \texttt{threedhst}
pipeline\footnote{\url{http://code.google.com/p/threedhst/}} \citep[][]{brammer:12a}. The individual
exposures of each visit were then combined into a mosaic by interlacing the
pixels to a subgrid of pixels that are exactly half their original size.
Interlacing pixels from
input images which have 0.5 pixel offsets (by design in 3D-HST and rounded for CANDELS-Primo) results in a one-to-one correspondence between input and output pixels
\citep[e.g., ][]{vandokkum:00} and offers the key benefit of
preserving the individual pixel errors.  

With the HUDF12 $H_{160}$ image defining the 1.6~$\mum$ flux densities and spatial
morphologies, we compute a full quantitative contamination model of all objects in the field to $H_{160}
< 28$.  The modeling technique, first presented in \cite{brammer:12b}, includes the first 4 spectral orders, among them the compact zeroth order that
can resemble emission lines.  The
models of bright objects with $H_{160} < 23$ are refined
based on their observed spectra, while fainter objects are assumed to have
continua flat in units of $f_\lambda$. Therefore, emission lines in fainter galaxies are not included in the contamination model.  

To generate a deep 2D spectrum from all UDF grism integrations, we
first extract two-dimensional spectra from each visit.  These spectra
are aligned in two dimensions to the nearest pixel in the
subsampled grid and the aligned pixels are then combined with inverse
variance weights including a term to down-weight contaminated
pixels.  Stacked spectra from the central 3
arcmin$^2$ of the UDF have a combined exposure time of 40.5~ks.

\section{A Tentative Emission Line in \UDFj}
\label{s:udfj}

\subsection{The Deep Near-IR Spectrum}

\begin{figure*}
\epsscale{1.1}
\plotone{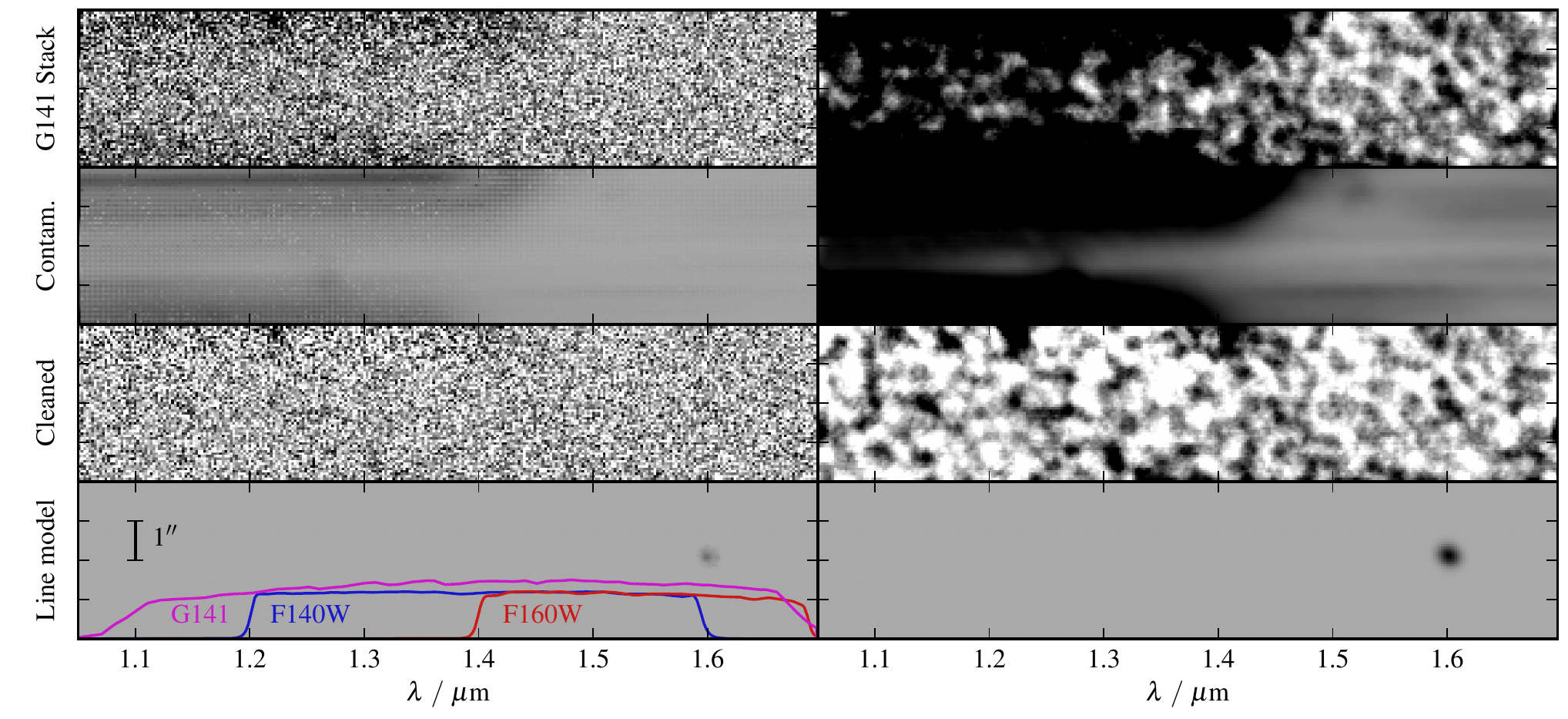} 
\caption{The deep G141 grism spectrum of \UDFj. The left panels show the pixels as
observed and modeled, and the right panels show the spectra cross-correlated with the $H_{160}$ object profile to enhance weak features. The first row shows the stacked
spectrum as observed. The second row shows the contamination model (\S\ref{s:observations}), with some aliasing due to the unequal
weights of the interlaced Primo mosaics.  The third row shows the observed
spectrum cleaned of this contamination.  There is a weak emission feature in the cross-correlation spectrum at $\lambda=1.599\,\mum$. The bottom
row shows the emission line model of \UDFj\, where the morphology is derived from the observed $H_{160}$ image (see also
Fig.\,\ref{f:oned_extraction}).\label{f:spec6001}}
\end{figure*}

The two-dimensional 40.5~ks G141 spectrum of \UDFj\ is shown in
Fig.\,\ref{f:spec6001}.  While some contamination remains
in the stacked spectrum, the orientations of
the UDF/G141 visits turn out to be nearly perfect for avoiding
significant contamination of the \UDFj\ spectrum, particularly at $\lambda >
1.4\,\mum$. The
contamination-subtracted spectrum is remarkably clean
with none of the systematics that typically plague ground-based NIR
spectroscopy \citep[see, e.g.,][]{lehnert:10}. The cluster of bright
pixels at the top of the spectrum is likely a faint contaminant emission line not
included in the model; it lies well away from any
contribution from \UDFj.

The ``salt-and-pepper'' appearance of the 2D spectrum is the result of the uncorrelated noise of
the individual pixels. The distribution of cleaned pixel values is gaussian
with $\sigma_\mathrm{obs}=0.0033$\,\epers. Random gaussian
deviates scaled by the propagated pixel errors have
$\sigma_\mathrm{err}=0.0036$\,\epers. This slight overestimation of the errors
by $10\%$
is significantly lower than the factors of
60--70\% that result from typical applications of the Drizzle algorithm
\citep{fruchter:09}.


\subsection{A $2\sigma$ Emission Line Detection at $\lambda=1.599\,\mum$}
\label{s:detection}


\begin{figure*}
\epsscale{1.1}
\plotone{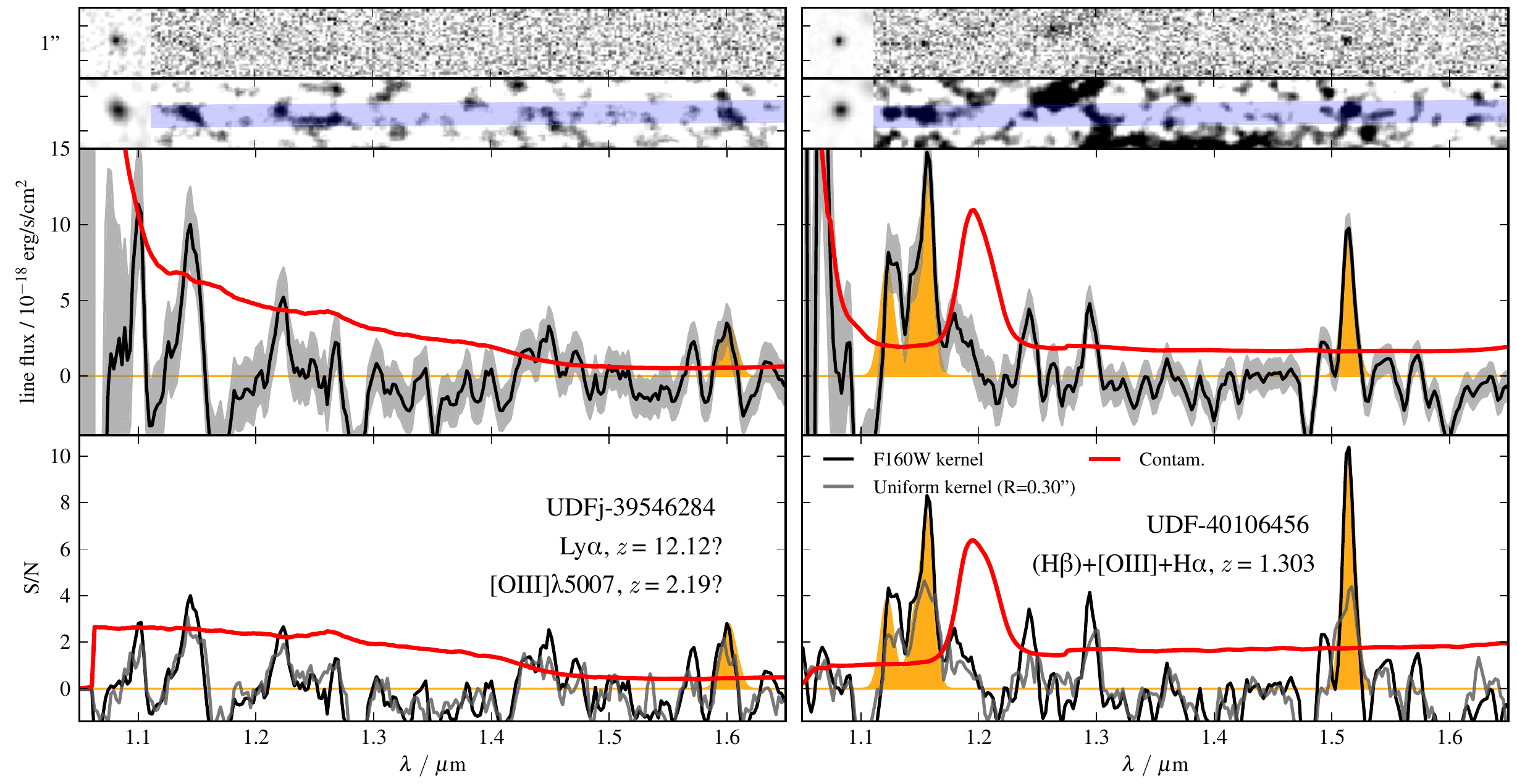} 
\caption{Optimal extraction of the \UDFj\ spectrum (left) and that of a comparison
object, \compareID (right). The top panels show the stacked 2D spectra and the
spectra cross-correlated with the $H_{160}$ image kernel, which is indicated at
left. The upper spectra show the 2D cross-correlation extracted
along the trace and scaled by the G141 sensitivity curve and $\Delta\lambda$
of the spectrum, which for a uniform kernel would give an integrated line flux
with the units as labeled. The same cross-correlation is applied to the 2D
pixel variance array, yielding the uncertainties shaded in gray. The bottom panels show the signal-to-noise spectrum.  While there are additional apparently significant features at $\lambda < 1.5\,\mum$, these are likely due to contamination rather than coming from \UDFj\ (\S\ref{s:reality}).  The gray curves in the bottom panels show the
S/N in $R=0\farcs3$ apertures along the trace. The red curves indicate the potential contamination from nearby
sources. The filled orange regions are model emission line spectra, whose shapes match those of the observed lines.
\label{f:oned_extraction}}  
\end{figure*}

Slitless \HST\ grism spectra offer a distinct advantage compared to
ground-based data for detecting weak emission line features: the 2D
spectrum provides an image of the galaxy at the wavelength of the emission line \citep[see, e.g.,][]{brammer:12b},
and the morphology of the galaxy is known a priori from the deep broad-band
WFC3 imaging. As long as the emission line morphology and the broad-band
morphology are similar (which is generally the case at $z\sim 1$;
\citealp{nelson:12}), or the line contributes significantly to the broad-band
flux (which is the case here; see below), we can search for features in the 2D
spectrum whose morphology corresponds to that of the broad-band image.

While there are no obvious emission lines in the raw 2D spectrum, there is a clump of pixels at $\sim 1.6\,\mu$m which is marginally enhanced compared to its
surroundings.  To improve the signal-to-noise of this feature, we cross-correlate the
2D spectrum with a kernel defined by the central $R=0\farcs3$ of the deep
$H_{160}$ thumbnail of \UDFj. This kernel is constructed by extracting kernels for each of the individual UDF/G141 visits with different orientations and combining them weighting by the median error in their corresponding 2D grism spectra. Thus,
the object profile is slightly smoothed but will reflect the 2D morphology
of lines in the stacked spectrum. The cross-correlation spectrum is shown in the
right-hand panels of Fig.\,\ref{f:spec6001}, and the strength of the feature at
1.6~$\mum$ is enhanced compared to the raw 2D spectrum.  We verified that the weak feature
is not visible in a similar analysis of any single UDF/G141 visit,
confirming that the feature does not arise from a flux excess in one of
them individually, such as a group of hot pixels or an un-flagged cosmic ray.

We show a one-dimensional spectrum extracted along the trace in
Fig.\,\ref{f:oned_extraction}. We compute an associated uncertainty at each
pixel along the trace, highlighted in blue, by cross-correlating the squared
error array with the same $H_{160}$ kernel. Furthermore, we compute a 2D model spectrum as
in Fig.\,\ref{f:spec6001} with known position and integrated line flux, and
extract its cross correlation spectrum in the same way as the observed
spectrum. The feature at $\lambda=1.599\pm0.004\,\mum$ is detected at 2.7$\sigma$ with
an integrated line flux of \lineflux.  The probability of finding a gaussian noise feature with equal or greater
significance is $\sim$10\% for 30 independent resolution elements  ($\sim$3000\AA).

\subsection{The Reality of the $1.599\,\mum$ Feature}
\label{s:reality}

There are regions in the 2D cross-correlation spectrum with
apparently similar significance to the feature at 1.6 $\mum$. Many of them can
be rejected because they do not fall precisely along the trace of the
grism spectrum.  There are enhancements found near the trace at
$\lambda=1.14\,\mum$ and $1.22\,\mum$, though these wavelengths suffer from
higher contamination perhaps consisting of faint emission lines not included in the contamination model.  The \HST/WFC3
broad-band photometry places strong constraints on the possibility that
these lines come from \UDFj: either of the bluer lines alone would result in
$Y_\mathrm{105}$ or $J_\mathrm{125}$ two magnitudes brighter than
the $2\sigma$ HUDF12 limits.

In contrast, the $1.599\,\mum$ line is consistent with the HUDF12
photometry. Within the same $0\farcs3$ aperture used to measure the line flux,
we measure $H_\mathrm{160}=29.04$ for the $H_{160}$ kernel. This is 27\% brighter
than the \cite{ellis:13} magnitude measured within an $0\farcs25$ aperture,
but is only 73\% of the total magnitude measured by \cite{bouwens:13}.  These differences are consistent with aperture corrections of the extended source, which shows a
faint extended tail to the NE in the $H_{160}$ image \citep{bouwens:13}. A pure emission line with the observed flux would result in
$H_\mathrm{160}=28.92$; the line accounts for $110\pm40$\%
(1$\sigma$) of the $H_{160}$ flux. This corresponds to a (1,2)$\sigma$ limit on the observed-frame
equivalent width of the line of $\mathrm{EW} > (7300,1300)$\,\AA. 


\begin{figure*}
\epsscale{1.1}
\plotone{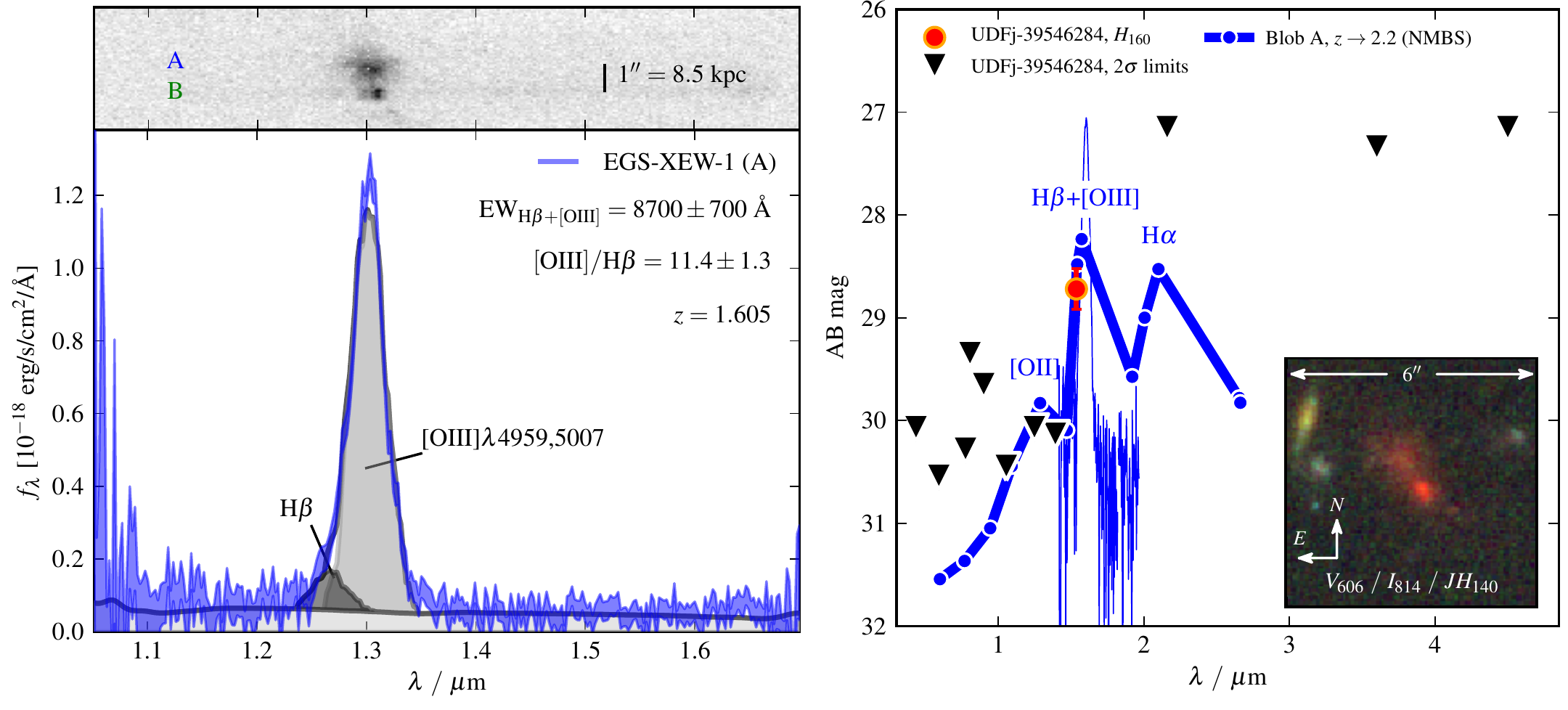} 
\caption{WFC3/G141 spectrum and broad-band SED of EGS-XEW-1, an ``[\ion{O}{3}] blob'' at
$z=1.605$ with extreme [\ion{O}{3}] emission lines and colors similar to \UDFj.  The asymmetric line profile of component A is consistent with lines of H$\beta$ and the \othree\ doublet as indicated; the lines are unambiguously resolved for component B, at the same redshift. The right panel shows the SED of the blob normalized to that of
the \UDFj\ $H_{160}$ magnitude. Non-detections for \UDFj\ are shown with 2$\sigma$
upper limits taken from \cite{bouwens:13}. Along with the G141 spectrum, ground-based photometry of the blob from the NEWIRM Medium Band Survey (NMBS) \citep{whitaker:11}
is shown in blue, redshifted such that the [\ion{O}{3}] line would have been observed at
$1.6\,\mum$ and scaled to account for the different NMBS-J3, $JH_{140}$, and $H_{160}$
passbands. The inset shows a color thumbnail of the blob created
from \HST\ ACS/$V_{606}$, ACS/$I_{814}$, and WFC3/$JH_{140}$ images. \label{f:egs_blob}}
\end{figure*}

We test our technique by analyzing faint lines of another UDF
galaxy with a secure redshift, \compareID (right panels of Fig.\,\ref{f:oned_extraction}). This galaxy has
$H_\mathrm{160}=27.4$ and unambiguous emission lines of H$\alpha$ and \othree\
at $z=1.303$ visible in the stacked 2D spectrum before
cross-correlation. Using the cross-correlation technique we find a significant
detection of H$\beta$ even though the line is barely visible in the original
spectrum. Two spurious features are detected at
1.24$\,\mu$m and $1.29\,\mu$m with comparable flux as the tentative line
in \UDFj. These are both clearly associated with residual contamination from
another source at the top of the 2D spectrum. The H$\alpha$ line flux of
\compareID\ is $9.5\pm1.0\times10^{-18}\,\fluxcgs$, just three times brighter
than the \UDFj\ emission line. The detection significance is
S/N$\sim$10 and the cross-correlation
technique enhances the detection significance by a factor of two
compared to simple photometry along the trace within an equivalent $0\farcs30$
aperture. This example demonstrates that the cross-correlation technique works
and is able to recover extremely faint emission lines.

%
%
\section{Line identification}
\label{s:line_id}

\subsection{Is \UDFj\ at $z=12?$}

If the galaxy is at $z\sim 12$, as is suggested based on its strong
photometric break between the $JH_{140}$ and $H_{160}$ filters \citep{ellis:13, bouwens:13}, the line---if real---can be identified as Ly-$\alpha$ redshifted to
$z=12.12$. The rest-frame equivalent width of Ly-$\alpha$ would
not be unreasonably high at $>
170$\,\AA\ (accounting for the $z=12$ Lyman break in $H_{160}$); such values can be reached in young, low metallicity
starbursts at high redshift \citep{schaerer:03}.  However, such strong Ly-$\alpha$ emission might be unexpected early in the
reionization epoch when the neutral fraction is high (\citealp{santos:04}; but see also \citealp{dijkstra:11}).  More to the point, if the line is real it can account for
most or all of the $H_{160}$ flux, meaning that the photometric break does not necessarily reflect a strong
continuum break and the photometric redshift is unreliable. Therefore the line could also be a longer-wavelength feature at much
lower redshift, a possibility discussed by
\cite{ellis:13}. 

\vspace{0.2cm}

\subsection{Low redshift solutions---[\ion{O}{3}] at $z=2.2$}

Although we cannot exclude other low redshift identifications
such as  [\ion{O}{2}] at $z=3.28$ or H$\alpha$ at $z=1.44$, a likely possibility
is that the line is \othree\ at $z=2.19\pm0.01$.
A population of extreme [\ion{O}{3}] emission line galaxies at
$z\sim1.7$ was recently identified by \cite{vanderwel:11} from their
significant line contribution to the $J_{125}$ photometry in the CANDELS
survey.
Additional galaxies with [\ion{O}{3}] rest-frame equivalent widths
reaching 2000\,\AA\ have been identified in WFC3 grism spectroscopy by
\cite{atek:11} and \cite{brammer:12b}.
The strongest emitters in \cite{vanderwel:11} have
$J_{125}-H_{160}\sim -1$, with [\ion{O}{3}] in the $J_{125}$ band.
While such colors can mimic those of high-redshift dropout galaxies \citep[][]{atek:11}, even more extreme equivalent widths at $\gtrsim$3000\,\AA\ (rest-frame) are required to reach $J_{125/140} - H_{160} > 1.4$ observed for \UDFj.

\subsection{Discovery of a Possible Analog: An Extreme [\ion{O}{3}] Emitter at $z=1.605$}



We have discovered a very bright example of such an extreme emission line
galaxy in G141 grism observations of the EGS field (GO-12547, PI: Cooper). This remarkable object, EGS-XEW-1, located at $\alpha$=14:17:58.2, $\delta$=+52:31:35 and shown in
Fig.\,\ref{f:egs_blob}, is an ``[\ion{O}{3}] blob'' at $z=1.605$ with
$JH_{140}=21.5$ and a complex morphology comprised of diffuse (A) and compact
(B) components separated by 1$\arcsec$. The exposure times and dithered pixel
sampling for the Cooper et~al.~grism program are nearly identical to those of
3D-HST, and the spectra were processed with the 3D-HST interlacing software as
described above. The diffuse component, which is extended over more
than 1$\arcsec$ (8.5 kpc), has a combined observed-frame equivalent width of
nearly 9000\,\AA\ for the blended H$\beta$ and \othree\ emission lines. The more
compact component has EW=$3000\pm160$\,\AA. 

The broad-band SED of EGS-XEW-1 is compared to that of \UDFj\ in the right
panel of Fig.\,\ref{f:egs_blob}. The spectral break that results from a line
such as the one observed in EGS-XEW-1 is just large enough to satisfy the
2$\sigma$ limit on the $JH_{140}-H_{160}$ dropout color of \UDFj. However, if
\UDFj\ had exactly the same spectrum as EGS-XEW-1, the
[\ion{O}{2}]$\lambda$3727 line would probably be detected in the bluer WFC3
HUDF12 photometry. Interestingly, EGS-XEW-1 has much redder UV colors than
$f_\lambda\sim\lambda^{-2}$ observed for the high equivalent width starbursts
at similar redshifts \citep{vanderwel:11, brammer:12b}, which, if also true
for \UDFj, could help explain the lack of $z=2$ Ly-$\alpha$ observed for that
source \citep{ellis:13}. While not a perfect match, EGS-XEW-1 provides a
directly-observed route to a plausible low-redshift interpretation for the
\UDFj\ photometry.


\section{Discussion and Conclusions}
\label{s:discussion_and_summary}

We have described a deep \HST\ grism spectrum of the candidate $z=12$ galaxy
\UDFj. Using 
the known emission line morphology to increase the line sensitivity of the
slitless grism spectrum, we detect a tentative emission line at
$\lambda=1.599\,\mum$ with flux \lineflux\ and
observed-frame equivalent width $>7300$\,\AA.

While the current observations do not conclusively forbid the $z=12$ interpretation of \UDFj, a number of independent factors conspire to suggest that the 2.7$\sigma$ line is in fact
real and that the true redshift of \UDFj\ is $z\sim2.2$. First, the line is observed
at just the right wavelength and flux to satisfy both the HUDF $H_{160}$
detection and the $>$1.4 mag dropout at $\lambda < 1.595\,\mum$ in the bluer
\HST\ bands. If the feature were observed just 4 pixels
(93\,\AA) bluer it would violate the constraints on $JH_{140}$ by a
magnitude. Second, \cite{bouwens:13} demonstrate that without a significant contribution of Ly-$\alpha$ to the $H_{160}$ flux, the implied
rest-frame UV
luminosity for $z\sim 12$ is
some 20$\times$ higher than would be expected from the
evolution of the luminosity function at $z=8$--10.
Third, recent deep spectroscopic surveys of photometric dropouts
find a decreasing fraction Ly-$\alpha$ emitters with increasing redshift when the universe was increasingly neutral \citep[][]{pentericci:11, caruana:12}.  Fourth, we show that spatially-extended objects with $\mathrm{EW} > 7300$\,\AA\ exist at $z\sim2$.  Finally, we find at
least two [\ion{O}{3}] emitters within 1000\,\kms\ of the $z=2.19$ solution for
\UDFj\ in the 3D-HST coverage of the full GOODS-South field, the nearest
separated by 69$\arcsec$ (570 kpc); the fact that other galaxies exist at this exact redshift increases
the probability that the line is real and is \othree.

If the physical properties of \UDFj\ are similar to either the [\ion{O}{3}] blob EGS-XEW-1 or to the high equivalent width
starbursting dwarf galaxies studied by \cite{vanderwel:11},
\UDFj\ at $z=2.2$ would represent a new class of object 750 times fainter
than the
former or $\sim$30 times fainter than the latter. Scaling from the typical
stellar masses of the \cite{vanderwel:11} galaxies, \UDFj\ would have a stellar mass of order
$10^6~M_\odot$, similar to the mass of a single massive star cluster
\citep[e.g.,][]{reines:08}. The extended $H_{160}$ morphology would then indicate
the distribution of ionized gas surrounding the cluster.

The most urgent question is whether the line is real and associated
with \UDFj\ as opposed to coming from a nearby contaminating object spectrum
or simply being a clump of positive noise fluctuations. If the line is real
and is [\ion{O}{3}] at $z=2.2$, it should be possible to detect the 4959 and
5007 \AA\ lines of the doublet separately, either in (much) deeper grism data
or in deep ground-based spectroscopy.  H$\alpha$ may also be visible in the $K$-band at a similar flux as [\ion{O}{3}] (Fig.\,\ref{f:egs_blob}).  Regardless of the reality of the emission line, we have shown
that the broad-band SED of \UDFj\ provides a reasonable match to a low redshift SED,
specifically the \textit{observed} spectrum of EGS-XEW-1, an ``[\ion{O}{3}] blob'' at
$z=1.605$, redshifted to $z=2.2$.
This study demonstrates that the G141 grism on WFC3 can
provide clean spectra
with well-understood noise properties in deep integrations up
to at least 40\,ks (17 orbits). 

%
%
\vspace{-0.3cm} 
\acknowledgements

\noindent We gratefully acknowledge funding support from the Marie Curie Actions of the European Commission (FP7-COFUND) and STScI
grant GO-12177.

{\it Facilities:} \facility{Hubble Space Telescope (WFC3)}


%
%

\end{document}